\documentclass[twocolumn,showpacs,nofootinbib]{revtex4}
\usepackage[utf-8]{inputenc}
\usepackage{amssymb,eucal,amsthm,epsf,graphics,float}
\usepackage[english]{babel}
\def\vf{v_{\!f}}
\begin{document}
\title{
Magnetic field induced singlet - triplet phase transition in quasi one-dimensional organic superconductors}
\author{N. Belmechri}
\affiliation{Laboratoire de Physique des Solides,
UMR 8502 CNRS – Université Paris-Sud 11}
\author{G. Abramovici}
\affiliation{Laboratoire de Physique des Solides,
UMR 8502 CNRS – Université Paris-Sud 11}
\author{M. Héritier}
\affiliation{Laboratoire de Physique des Solides,
UMR 8502 CNRS – Université Paris-Sud 11}
\author{S. Haddad}
\affiliation{
Laboratoire de Physique de la matière condensée, Faculté des Sciences de Tunis}
\author{S. Kaddour}
\affiliation{
Laboratoire de Physique de la matière condensée, Faculté des Sciences de Tunis}

\begin{abstract}
We propose a theoretical model of quasi-one-dimensional superconductors, with attractive electron-electron interactions dominant in the singlet $d$-wave channel and sub-dominant in the $p$-wave channel. We discuss, in the mean field approximation, the effect of a magnetic field applied perpendicularly to the direction of the lowest conductivity. The lowest free energy phase corresponds to a singlet $d$-wave symmetry in low fields, but to a triplet symmetry in high fields. A first order singlet-triplet phase transition is expected at moderate applied fields of a few teslas. We propose to ascribe the recent critical field and NMR experimental data, observed in superconducting $(TMTSF)_2ClO_4$ to such an effect.
\end{abstract}
\maketitle

The nature of superconductivity in the family of the quasi-1D organic superconductors~\cite{mazaud,jerome,jerome2} $(TMTSF)_2X$ ($X=PF_6$, $ClO_4$,...) has been a long standing issue for the last three decades. There is still a debate whether it is a conventional superconductivity, with a completely gapped Fermi surface, or an unconventional one with points or lines of nodes on the gap. Many experiments have tried to address this question using different techniques such as NMR relaxation rate, thermal conductivity, non magnetic impurity effect on $T_c$, Knight Shift and upper critical field measurements.

The temperature dependence of the proton spin lattice relaxation rate, measured by Takigawa \textit{et al.}~\cite{takigawa} in the $ClO_4$ compounds, suggested the presence of lines of nodes in the gap on the Fermi surface. On the other hand, Belin \textit{et al.}~\cite{belin} measurements of the temperature dependence of the thermal conductivity argued for a fully gapped Fermi surface. More recently, Joo \textit{et al.}~\cite{joo} studied the effect of non magnetic impurities on the superconducting critical temperature $T_c$~\cite{joo2}, and showed unequivocally that the order parameter changes sign on the Fermi surface and that consequently the gap has nodes where it is zero on the Fermi surface. This is in complete agreement with a model of unconventional superconductivity in these organic materials. 

 However, whether the pairing of the electrons in the superconducting state is in the singlet or the triplet symmetry remains an unsolved question. Measurements of the upper critical field of superconductivity in these materials~\cite{lee,oh,chaikin} showed a superconducting state surviving up to a field as high as 9 T in the case of the $PF_6$ compounds~\cite{lee}, and at least 5 T in the case of the $ClO_4$ compounds~\cite{oh}. These fields exceed by far the Clogston-Pauli paramagnetic limit~\cite{clogston} for homogeneous singlet superconductivity, which is estimated to be of the order of 1.84 $T_c$, with a $T_c$ around 1.1 K in these materials. This result is a strong indication that at high magnetic fields superconductivity could not be of a homogeneous singlet type. 
Many theories~\cite{lebed,lebed2,maki} have been proposed to explain this behaviour by a triplet superconducting state, which is a non Pauli limited state. At low magnetic fields, however, besides the observation by Andres \textit{et al.}~\cite{anders} of a diamagnetic signal when the field is perpendicular to the chains indicating a singlet state, measurements of the Knight-Shift~\cite{shinagawa} showed a significant variation from the normal state behaviour of the spin susceptibility of Cooper pairs which decreased with decreasing temperature in the two directions of the $a{-}b'$ planes. This behaviour was interpreted as being the signature of a singlet state at low fields. Therefore, it seems that there is a contradiction between a low magnetic field singlet behaviour and a high magnetic field non singlet one.

Other authors, however~\cite{dupuis,shimahara}, pointed out that a singlet superconducting state may survive the Pauli paramagnetic limit by allowing for a spatially modulated order parameter, forming an inhomogeneous singlet state, or FFLO state~\cite{fulde,larkin,maki2}, where the two electrons of the Cooper pair will have different wave vectors to compensate for the Zeeman splitting.
Maki \textit{et al.} was the first to consider FFLO in $d$-wave superconductors
in $CeCoIn_5$ and to consider the orbital effect\cite{won,miclea}.
Nevertheless, a phase transition to an inhomogeneous FFLO state has not yet been identified experimentally with certainty in Bechgaard salts. 

Another important property of these quasi-1D materials, which must not be ignored to explain their behaviour in magnetic field, is that under an increasing magnetic field parallel to the $b'$ direction the electron motion gets confined to the $a{-}b$ planes of the crystalline lattice%
: when the magnetic field is large enough ($H\gtrsim2$T), the electronic orbitals
in each $a{-}b$ plane are decoupled. Therefore the orbital pair breaking effect
is almost suppressed and can be neglected in that case\cite{lebed3,lee2}.
Using this particular characteristic, we study a simple model which considers the possibility of the phase transition between a singlet superconducting state at low magnetic fields, and a triplet one at higher fields, in order to explain the discrepancy between the low field singlet behaviour and the high field triplet one.

The purpose of this article is to describe this transition and not to
give the most accurate and most realistic values of the critical fields $H_{c2}(T)$,
as we have neglected the orbital pair breaking effect. Nevertheless, it is true
that this orbital effect is much weaker in Bechgaard salts than in isotropic
metals, because of their lamellar character, as discussed below.

Eventually, the transition between singlet and triplet superconductivity
results from the small difference between the corresponding values of the
renormalized scattering coefficients $g_s$ and $g_t$, as they have been
calculated by a renormalisation group method\cite{abramovici,abramovici2}.
This implies indeed a small free energy gap between both configurations.

\paragraph*{Theoretical model}

One of the most important features of the $(TMTSF)_2X$ compounds is the very weak electron transfer in the $c$ direction transverse to the $a{-}b$ planes~\cite{jerome,ishiguro}. The transfer integral in this direction is $t_c = 10$ K, which is an order of magnitude less than that in the $b$ direction, with $t_b = 300$ K, and two orders of magnitude less than that in the chains direction $a$, with $t_a = 3000$ K. The particular structural feature of these compounds gives them a lamellar character, and hence reduces considerably the orbital destructive effect against superconductivity. Furthermore, when the magnetic field is oriented parallel to the $b'$ direction at low temperatures, the semi-classical electron motion can be shown to be oscillatory in the $c$ direction~\cite{yonezawa} with the amplitude 
$$
\delta z = {4 t_c \over e\, \vf  H}
 $$
where $\vf$ is the Fermi velocity, $e$ the electron charge and $H$ the magnetic field.

Such a mechanism is somewhat analogous to the field-induced one-dimensionalisation leading to the FISDW~\cite{gorkov,heritier}. The electron system, in fact, undergoes a dimensional crossover from a low field 3D regime to a higher field effective 2D regime, where the electron motion is confined to the $a{-}b$ planes and the interlayer hopping is considerably reduced. This phenomenon has first been predicted by Lebed~\cite{lebed3} based on a quantum mechanical calculation taking into account the highly anisotropic shape of the Fermi surface. Magnetoresistance measurements along the $c$ direction~\cite{danner,behnia} have confirmed these predictions. This confinement phenomenon reduces the screening currents in the $c$ direction and therefore cancels out their destructive effect on superconductivity. In our model, we take these considerations into account by assuming that the orbital pair breaking effect of the magnetic field can be ignored as a first approximation. Therefore, the only pair breaking effect of the magnetic field that will be taken in our case is the Zeeman splitting of the electron energy. The inter-plane transfer, $t_c$, is implicit, so that the mean field approximation is justified. However, it is weak enough to be ignored in the subsequent calculations.

We consider a mean field Hamiltonian which allows for both singlet and triplet order parameters. At high magnetic fields the electron spin susceptibility is that of the normal state. This is why we assume that  the field induced triplet state is the equal spin pairing state (ESP)~\cite{balian}, for it has no $S_z = 0$ component and therefore has the same spin susceptibility as the normal state. We will consider it throughout this paper. Furthermore we assume a unitary state with real order parameters. The singlet and triplet order parameters are assumed to have $d$-wave and $p$-wave orbital symmetry respectively, see equations (5) and (6). The FFLO case, where the mixed singlet and triplet order parameters are spatially modulated, was previously studied by Shimahara~\cite{shimahara2,shimahara3}. In our model we will only consider the homogeneous state. A state in which singlet and triplet symmetry coexist does not seem favourable in our model.

The mean field Hamiltonian is described as follows
\begin{equation}
H=H_0+H_s+H_t	
\end{equation}

\begin{equation}
H_0=\sum_{\vec k \sigma} \xi_{\vec k \sigma} c_{\vec k \sigma}^\dag  c_{\vec k \sigma}
\end{equation}

\begin{equation}
H_s=\sum_{\vec k \sigma} \Delta_s (\vec k) \{c_{-\vec k \downarrow } c_{\vec k \uparrow } + c_{\vec k \uparrow }^\dag c_{-\vec k \downarrow }^\dag \}
\end{equation}

\begin{equation}
H_t={1 \over 2}\sum_{\vec k} \Delta_t (\vec k) \{c_{\vec k \downarrow } c_{-\vec k \downarrow } + c_{-\vec k \uparrow } c_{\vec k \uparrow } + h.c. \}
\end{equation}
with the following definitions for the order parameters, where $s$
and $t$ subscripts stand for singlet and triplet respectively,
\begin{eqnarray}
\Delta_s (\vec k)&=&\Delta_{s,\uparrow \downarrow} (\vec k)\\
&=&\sum_{\vec k'} V_s (\hat k, \hat k') \{ \langle\langle c_{-\vec k' \downarrow } c_{\vec k' \uparrow }\rangle\rangle - \langle\langle c_{-\vec k' \uparrow } c_{\vec k' \downarrow }\rangle\rangle \}\nonumber\\
\Delta_t (\vec k)&=&\Delta_{t,\uparrow \uparrow} (\vec k)
=-\Delta_{t,\downarrow \downarrow} (\vec k)\nonumber\\
&=&2 \sum_{\vec k'} V_t (\hat k, \hat k') \langle\langle c_{-\vec k' \uparrow} c_{\vec k' \uparrow }\rangle\rangle
\end{eqnarray}
with
\begin{eqnarray}
V_s(\hat k,\hat k')&=&V_s^0 \left(2 (\hat k . \hat k')^2 - 1\right)\\
V_t(\hat k,\hat k')&=&V_t^0 \hat k . \hat k'
\end{eqnarray}

In our model, we use the following assumptions~:
\begin{eqnarray}
\Delta_s (\vec k)&=&\Delta_s^0 (2 \hat k_x^2 -1)\\
\Delta_t (\vec k)&=&\Delta_t^0\,\hat k_x\quad
\hbox{and}\quad\Delta_{t,\uparrow \downarrow} (\vec k)= 0
\end{eqnarray}

$V_s(\hat k,\hat k')$ and $V_t(\hat k,\hat k')$ are respectively the $d$-wave and $p$-wave parts of the interaction potential, which is assumed to be attractive in both parts. The possibility of coexistence of these two attractive parts of the interaction has been previously discussed by many authors~\cite{shimahara3,kuroki}. In the g-ology model~\cite{jerome,solyom} both interactions are included in the Hamiltonian and their relative relevance determines the symmetry of the superconducting state. In our model, we do not discuss the physical origin of these interactions, whether they are mediated by phonons or by magnetic fluctuations, or any other mechanisms. We assume the coupling constants, $V_s$ and $V_t$, to be temperature and magnetic field independent. On the other hand, in the case of a very strong magnetic field, the coupling constant may indeed depend on the field strength. This case, which has been discussed by Kuroki et al.~\cite{tanaka}, will not be discussed here.

As a first approximation we will consider an isotropic dispersion relation. We think that the anisotropic aspect of the dispersion relation will not change qualitatively our results, as will be reported elsewhere. In the following, we will study the relative stability of a $d$-singlet and a $p$-triplet superconducting state. We are interested, in particular, in the possibility of a phase transition between these two states in an increasing magnetic field. 

\paragraph*{Self consistent equations}

It is not difficult to diagonalize the model Hamiltonian eq.~(1). Here, we use the thermal Green's functions to solve it. The equations of motion for the relevant Green's functions yield the following equations in matrix form:
\begin{widetext}
\begin{equation}
\pmatrix{i \omega_n -\xi_{\vec k \uparrow}&-\Delta_{s, \vec k}&-\Delta_{t, \vec k}\cr -\Delta_{s, \vec k}&i \omega_n +\xi_{\vec k \downarrow}&0\cr -\Delta_{t, \vec k}&0&i \omega_n +\xi_{\vec k \uparrow}\cr}
\pmatrix{-\langle\langle c_{\vec k \uparrow} c_{\vec k \uparrow}^\dag\rangle\rangle\cr\langle\langle c_{\vec k \uparrow}^\dag c_{- \vec k \downarrow}^\dag\rangle\rangle\cr\langle\langle c_{\vec k \uparrow}^\dag c_{- \vec k \uparrow}^\dag\rangle\rangle\cr}
=\pmatrix{1\cr0\cr0\cr}
\end{equation}
\end{widetext}
where $\omega_n = (2n+1) \pi T$. Solving eq.~(9) for the relevant Green's functions, $\langle\langle c_{\vec k \uparrow}^\dag c_{- \vec k \downarrow}^\dag\rangle\rangle$ and $\langle\langle c_{\vec k \uparrow}^\dag c_{- \vec k \uparrow}^\dag\rangle\rangle$, and performing the frequency sum we can write the self consistent equations for the order parameters of equations~(5) and (6) as
\begin{eqnarray}
\Delta_{s,\vec k}&=&\sum_{\vec k'} { V_s (\hat k, \hat k') \Delta_{s,\vec k'} \over 2 E_{c,\vec k'}  } \{  (\Delta_{t,\vec k'}^2+(\mu_e H)^2-E_{c,\vec k'})\;thc_-\nonumber\\
&&- (\Delta_{t,\vec k'}^2+(\mu_e H)^2+E_{c,\vec k'})\;thc_+\}\\
\Delta_{t,\vec k}&=& -\sum_{\vec k'} { V_t (\hat k, \hat k') \Delta_{t,\vec k'} \over 2 E_{c,\vec k'}} \{ (E_{c,\vec k'}-\Delta_{s,\vec k'}^2+\mu_e H \xi)\;thc_-\nonumber\\
&&+ (E_{c,\vec k'}+\Delta_{s,\vec k'}^2-\mu_e H \xi)\;thc_+\}
\end{eqnarray}
where $\mu_e$ is the electron magnetic field and we have defined
\begin{eqnarray}
E_{o,\vec k}^2&=&\xi_{\vec k}^2+\Delta_{s,\vec k}^2\nonumber\\
E_{\vec k}^2&=&\xi_{\vec k}^2+\Delta_{s,\vec k}^2+\Delta_{t,\vec k}^2\\
E_{c,\vec k}^2&=&(\mu_e H)^2 E_{o,\vec k}^2 + \Delta_{s,\vec k}^2 \Delta_{t,\vec k}^2\nonumber
\end{eqnarray}

\begin{equation}
thc_\pm={\tanh({\beta \over 2} \sqrt{E_{\vec k'}^2 +(\mu_e H)^2\pm2E_{c,\vec k'}}) \over \sqrt{E_{\vec k'}^2 +(\mu_e H)^2\pm{c,\vec k'}}}
\end{equation}

It is straightforward to see that if we put $\Delta_t = 0$ in equation~(12) the field and temperature dependence of $\Delta_s(T,H)$ is given by
\begin{eqnarray}
\Delta_{s,\vec k}&=& -\sum_{\vec k'} { V_s (\hat k, \hat k') \Delta_{s,\vec k'} \over 2 E_{o,\vec k'}  }
\{\tanh({\beta\over2} (E_{o,\vec k'}- \mu_e H))\nonumber\\
&&+\tanh({\beta\over2} (E_{o,\vec k'}+ \mu_e H)) \}
\end{eqnarray}
which is nothing but the BCS gap equation with the magnetic field acting only on the spins of the electrons. When $\Delta_s=0$ in equation (13) we get another correct limiting case, that is,
\begin{equation}
\Delta_{t,\vec k}= -\sum_{\vec k'} V_t(\hat k, \hat k') \Delta_{t,\vec k'} {\tanh({\beta\over2} \sqrt{(\xi_{\vec k'}-\mu_e H)^2+\Delta_{t,\vec k'}^2}~) \over \sqrt{(\xi_{\vec k'}-\mu_e H)^2+\Delta_{t,\vec k'}^2}  }
\end{equation}

This determines the field and temperature dependence of the $p$-triplet order parameter. 

By numerically solving equations (16) and (17) for $\Delta_s^0$ and
$\Delta_t^0$ with the $d$ and $p$ symmetries given respectively by equations~(9)
and (10), we could calculate the critical field of superconductivity in these two limiting cases, using the free energy difference with the normal state given by
\begin{equation}
\delta F=\int_0^{V^0}{dv^0 \over (v^0)^2}(\Delta^0\left(v^0)\right)^2
\end{equation}

Using equation (18) we could also calculate the critical field beyond which
the $d$-singlet state becomes energetically less favourable that the $p$-triplet one. In fig.~1 are shown the phase diagrams of the pure $d$-singlet and $p$-triplet cases, with the transition line between these two phases. The strength of the triplet interaction, and accordingly the triplet zero-field critical
temperature, are unknown. Therefore we take the triplet critical temperature
as our free adjustable parameter. What should also be noted is that this transition to the triplet phase is a first order transition. However, accurate enough experimental measurements of thermodynamic quantities, such as specific heat or magnetic susceptibilities, have yet to be done to address the existence or the order of such a phase transition.

\begin{figure}
\epsfysize=20cm
\rotatebox{-90}{\scalebox{0.35}{\includegraphics{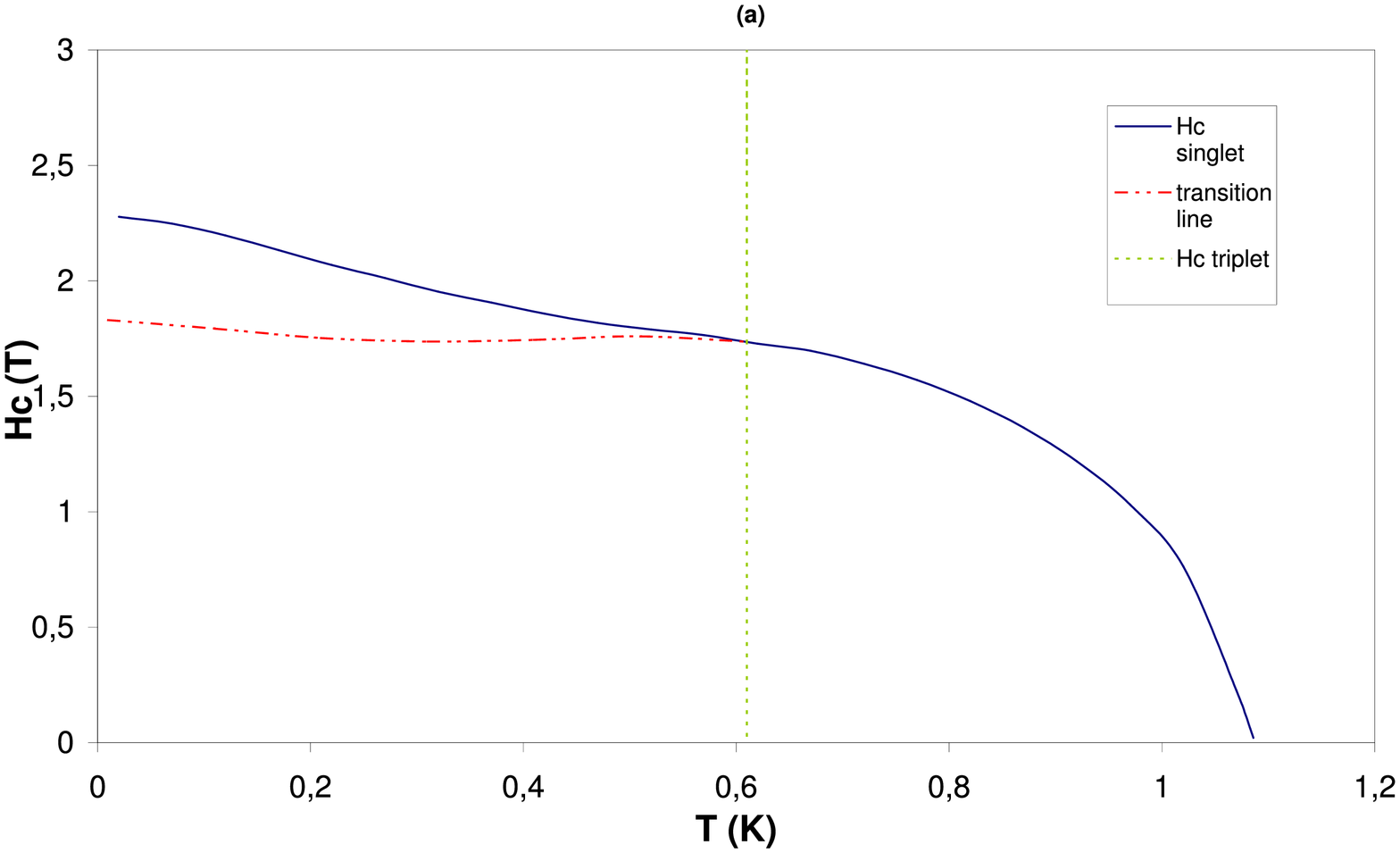}}}
\rotatebox{-90}{\scalebox{0.35}{\includegraphics{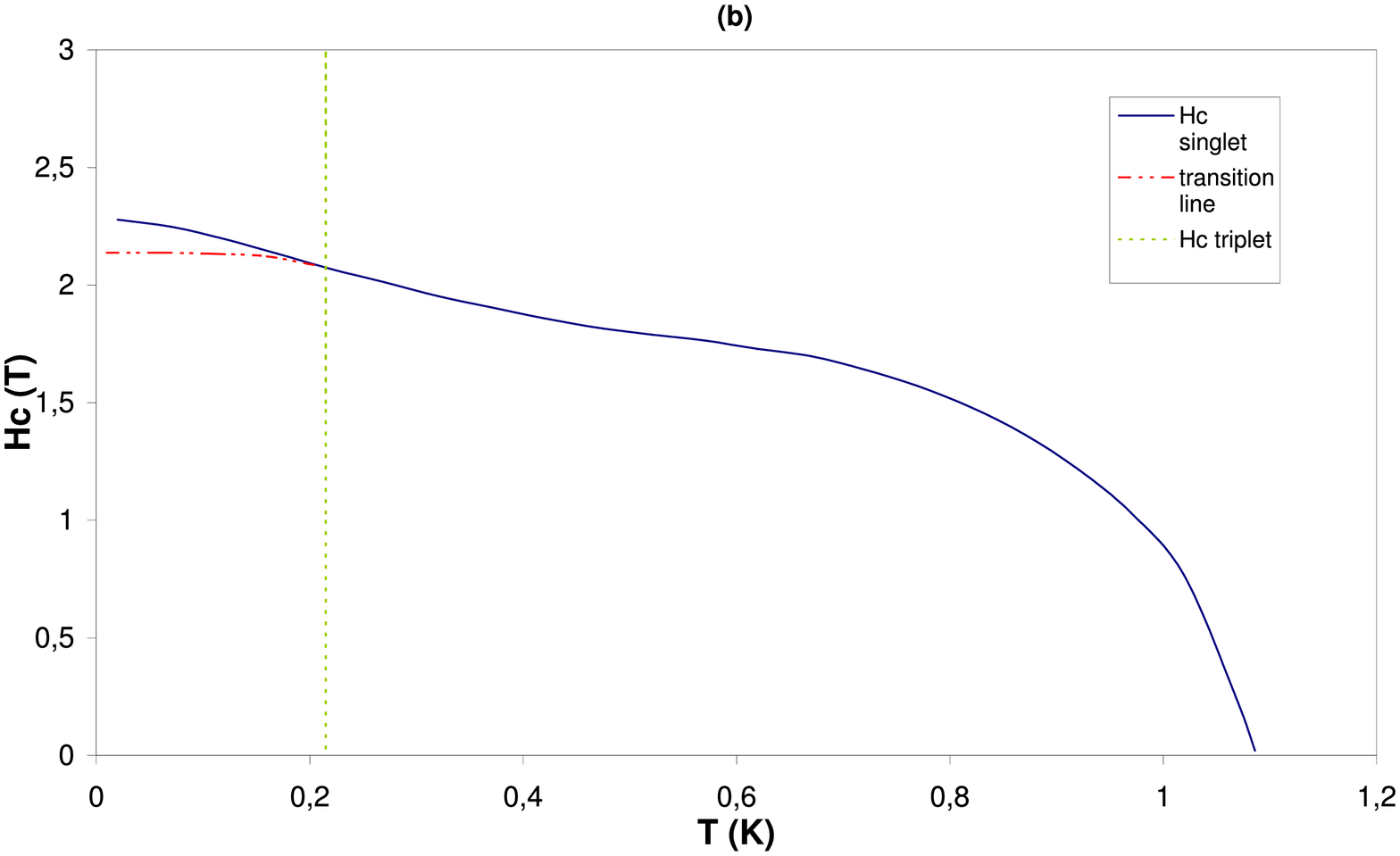}}}
\caption{Critical fields of superconductivity in the $d$-singlet (solid line) and $p$-triplet (dotted line) case, with a triplet critical temperature of (a)~0.62 K and (b)~0.22 K. The dash-dotted line is the transition line between the two phases. The infinite slope of the triplet linte is unrealistic, and a
consequence of the neglectance of orbital pair breaking effect.}
\end{figure}

Let us now study the possibility of a coexistence phase, where both the singlet and triplet order parameters are simultaneously non zero. When solving numerically the two coupled equations (12) and (13), we could not find any physically plausible solutions, which makes the coexistence state very unlikely to happen. To confirm this result we calculated a Ginzburg-Landau free energy development to the fourth order in $\Delta_s$ and $\Delta_t$, see eq.~(19). In the diagrammatic expansion of the free energy there are only three non-zero diagrams contributing to the ${\Delta_s}^2{\Delta_t}^2$ term. They are shown in fig.~2. All three diagrams have a positive sign in the $T{-}H$ plane, which makes coefficient $c$ in eq.~(19) positive for all values of $T$ and $H$, making the mixing term energetically unfavourable.
\begin{equation}
\delta F_c=\delta F_0 + a \Delta_s^2 + a' \Delta_t^2 + b \Delta_s^4 + b' \Delta_t^4 + c \Delta_s^2 \Delta_t^2
\end{equation}
with $\delta F_c$ being the free energy difference of the coexistence phase to
the normal state.

\begin{figure*}[t]
\epsfysize=8cm
\scalebox{0.3}{\includegraphics{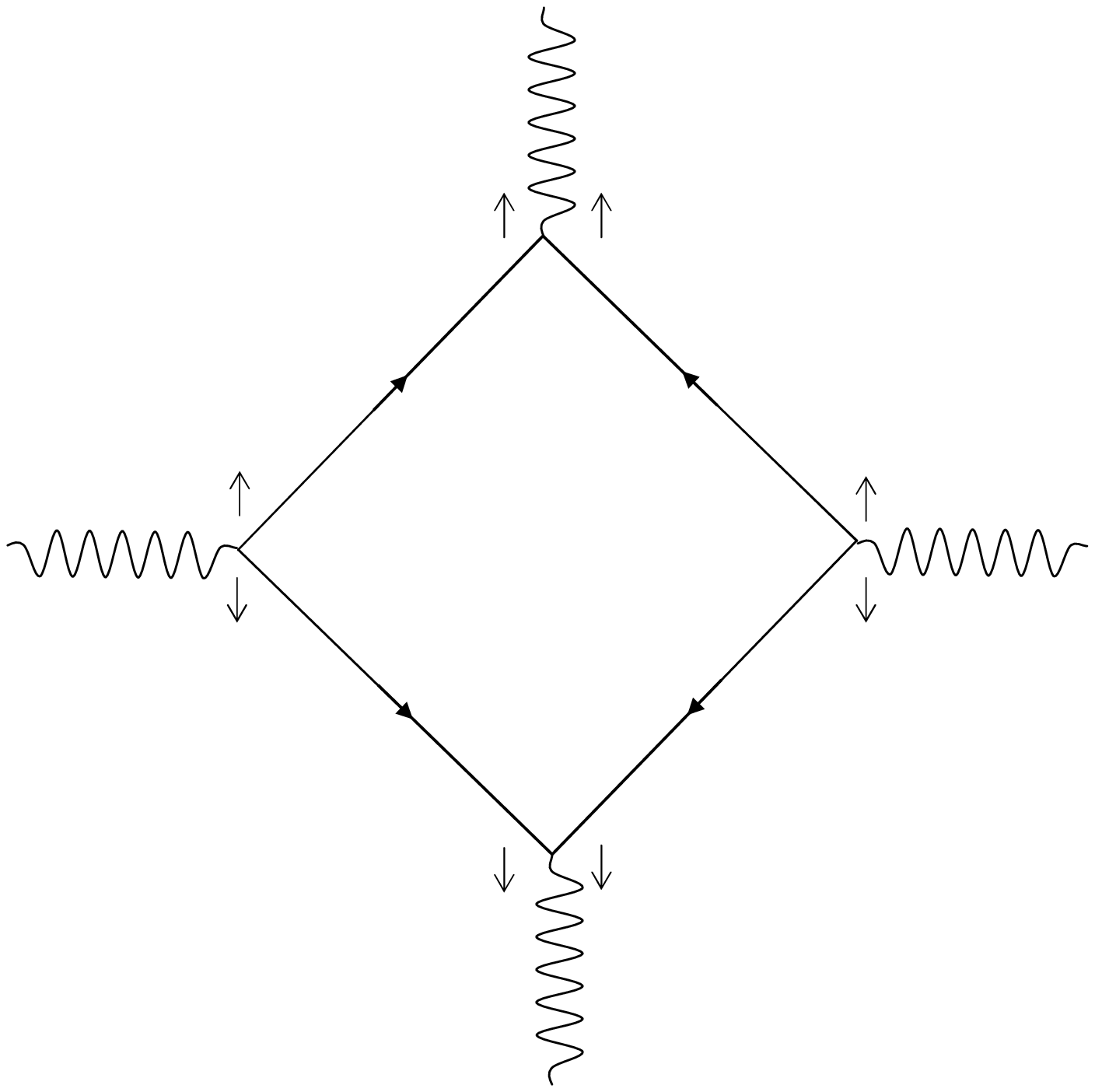}}
\scalebox{0.3}{\includegraphics{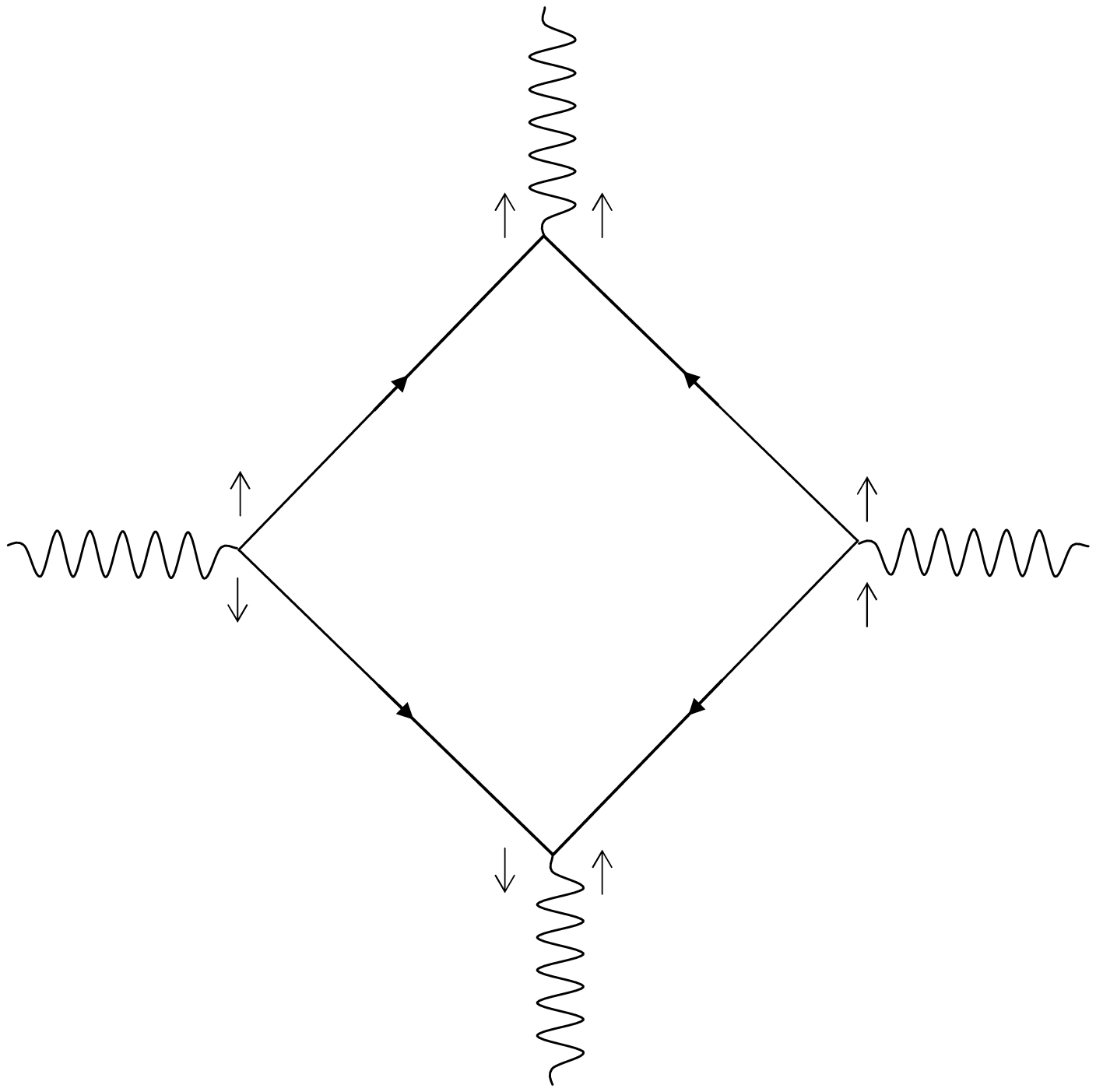}}
\scalebox{0.3}{\includegraphics{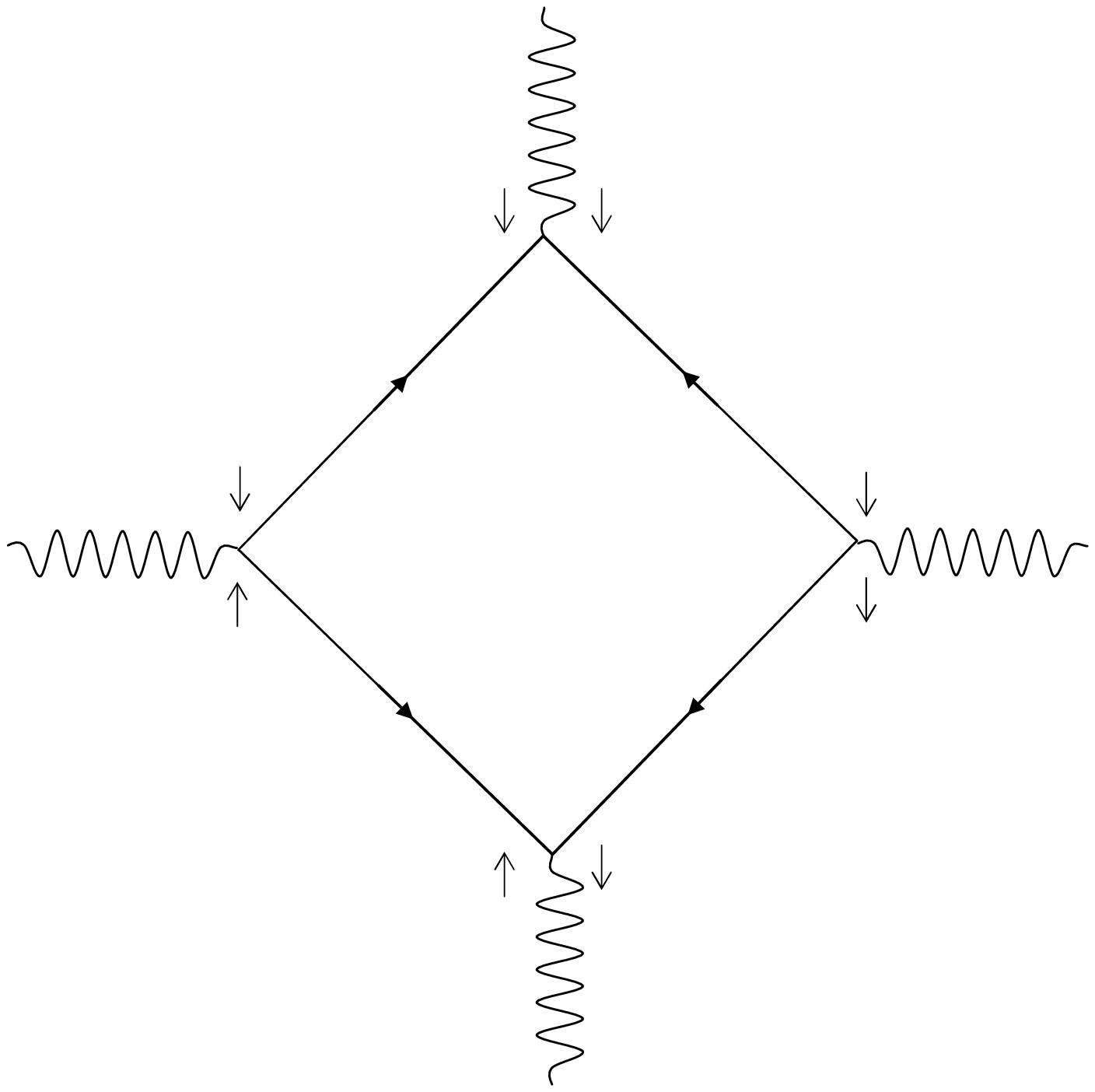}}
\caption{Diagrammatic expansion of the ${\Delta_s}^2{\Delta_t}^2$ term in the free energy.}
\end{figure*}

\paragraph*{Discussion}

In conclusion, we have presented a theoretical model in which we assume the simultaneous existence of singlet and triplet pairing interactions. This model, treated in a mean field approximation, shows a competition between two homogeneous superconducting phases, a $d$-wave singlet phase, stable in low applied field and a $p$-wave triplet superconducting phase, stable under large applied field. We have shown the existence of a phase transition from the $d$-singlet phase to the $p$-triplet phase in a moderate field, of the order of two teslas. A phase in which coexist these two symmetries is not stable in our simple model.
	
In this model, we therefore predict a very large critical field, much larger than the Clogston-Pauli limit. We are also able to understand the very peculiar behaviour of the NMR data recently observed in $(TMTSF)_2ClO_4$~\cite{shinagawa}: a usual Meissner effect is expected in the singlet  phase, while one should recover the normal phase susceptibility in the Equal Spin Pairing triplet phase, as experimentally observed.

A qualitative property concerning the anisotropy of the critical field can be easily predicted. As the applied magnetic field is rotated in the $a{-}b$ plane, the field orientation corresponding to the highest critical field should depend on the field strength. In low fields, typically $H$ smaller than about 2 teslas, the orbital effect of the field is too weak to induce the two-dimensional confinement observed at larger field~\cite{yonezawa}. Therefore, in this low field limit, we expect that the orientation of the highest critical field, which, of course is determined by the orbital effect is along the $a$-direction, in which the electron transfer integral is the largest. On the contrary, in the large field limit, the field induced two-dimensional confinement strongly reduces the electron transfer in directions perpendicular to $b'$. We therefore expect, as the field is increased above about 2 teslas, that the orientation of the maximum critical field is progressively rotated from the $a$-direction to the $b'$-direction.

In the large magnetic field case, the plane $a{-}b$ are decoupled and the orbital
pair breaking effect is drastically reduced and can be neglected, as we have
already discussed. In the small magnetic field case, however, we expect that
the remaining orbital effect will be similar for the singlet and for the
triplet cases, so that the difference between the triplet and the singlet
free energy is not noticeably modified; therefore we hope that the orbital
corrections will not destroy this transition. Moreover, we observe that
experimental values of the critical field $H_{c2}$ are noticeably increased
when the field is aligned along the $b'$ direction; this effect is coherent
with our assumption of a two-dimensional nature of the singlet/triplet
transition.

We have examined the case of a $p$-wave triplet state. We could have extended our study to an $f$-wave triplet state, which might be favoured by two-dimensional fluctuations~\cite{nickel,abramovici,abramovici2,maki}. We believe however
that similar results would have been obtained.

Our two dimensional model, described in eq. (1-4), can be obtained by a three
cut-off renormalisation group approach~\cite{montambaux}, in which the three
cut-offs are the bandwidth $E_0\approx6000$K, the cross-over temperature
$T_{\rm co}\approx100$K and the Zeeman energy $\mu_eH\approx2$K, and have different
orders of magnitude. In the range where $\omega<T_{\rm co}$, our Fermi liquid
analysis is valid. Since $\mu_eH<T_{\rm co}$, the Zeeman energy can also be
discussed in a mean field approach, as we did here.

Many authors~\cite{shimahara,shimahara2,shimahara3,suginishi} have proposed to interpret the experimental data by the field induced stabilisation, not of a triplet phase, but of a singlet-FFLO phase. Both phases have somewhat similar properties~: they appear in a first order transition ; they are very sensitive to non magnetic disorder, which might be the case of the high field phase ; they have a critical field which exceeds the Clogston-Pauli limit. However, there is a second paramagnetic limiting field beyond which the FFLO inhomogeneous singlet superconducting state will also be destroyed. This new paramagnetic limit was calculated by Lebed~\cite{lebed} in the quasi-1D $(TMTSF)_2X$ superconductors along the $b'$ direction, the direction of the highest critical field, and was estimated to be given by $H_{p\,\rm FFLO} = 0.6(t_a/t_b)^{1/2} H_p$ , where $H_p$ is the usual Pauli-limited critical field, $t_a$ and $t_b$ are the electronic transfer integral along $a$ and $b$. In the Bechgaard salts $H_{p\,\rm FFLO}$ should be of the order of 4 T. This limit was calculated without considering the orbital pair-breaking mechanism, which should reduce it further. It is not obvious to us that such a theoretical estimation of the FFLO critical field is large enough to account for the experimental values.

It seems to us that more experimental work is still necessary to discriminate between the various theoretical models proposed to interpret the experimental data

We gratefully acknowledge helpful and stimulating discussions with J. Friedel, D. Jérôme, and C. Pasquier. We are thankful for the french-tunisian cooperation CMCU (project 04 G1307).

\end{document}